\documentclass{PoS}
\usepackage{caption}
\usepackage{epsfig,amsmath}

\title{TMD parton distributions from parton showers}

\ShortTitle{TMDs from parton showers}

\author{\speaker{Melanie Schmitz}\\
        DESY, Hamburg, FRG\\
        E-mail: \email{melanie.schmitz@desy.de}}

\author{Francesco Hautmann\\
        Elementary Particle Physics, University of Antwerp, B 2020 Antwerp\\
        RAL, Chilton OX11 0QX and University of Oxford, Oxford OX1 3NP\\
        UPV/EHU, University of the Basque Country, E48080 Bilbao\\
        E-mail: \email{hautmann@thphys.ox.ac.uk}}

\author{Hannes Jung\\
        DESY, Hamburg, FRG\\
        E-mail: \email{hannes.jung@desy.de}}
        
\author{Sara Taheri Monfared\\
        DESY, Hamburg, FRG\\
        E-mail: \email{taheri@mail.desy.de}}

\abstract{We present the determination of Transverse Momentum Dependent (TMD) parton distributions from Monte Carlo parton showers. We investigate the effective TMD distributions obtained from the \textsc{pythia8} and \textsc{herwig6} parton showers and compare them to the TMD distributions determined within the Parton Branching method.}

\FullConference{XXVII International Workshop on Deep-Inelastic Scattering and Related Subjects - DIS2019\\
		8-12 April, 2019\\
		Torino, Italy}

\begin{document}

\section{Introduction}
The Parton Branching (PB) method is introduced in Refs \cite{HautmannJungLelek,HautmannJungLelek2,MartinezConnorHautmann} and gives an iterative solution for the evolution of both collinear and transverse momentum dependent parton distributions~\cite{Angeles-Martinez}. The advantage of this method is that the kinematics at every splitting process can be treated exactly since the solution is fully exclusive. It allows one to determine transverse momentum dependent PDFs (TMD).\\
In this report the determination of TMD parton densities from general parton shower event generators as \textsc{pythia8} \cite{PYTHIA} and \textsc{herwig6} \cite{HERWIG6,HERWIGgeneral} is presented.
\section{Parton Branching method}
The PB method gives the evolution equation for the momentum-weighted TMD parton density $\tilde{\mathcal{A}}_a=x\mathcal{A}_a$, 
\begin{align}
\tilde{\mathcal{A}}_a\left(x,\textbf{k},\mu^2\right)=&\Delta_a\left(\mu^2\right)\tilde{\mathcal{A}}_a\left(x,\textbf{k},\mu_0^2\right)+\sum_b\int\frac{\text{d}^2\textbf{q}^{\prime}}{\pi\textbf{q}^{\prime 2}}\frac{\Delta_a\left(\mu^2\right)}{\Delta_a\left(\textbf{q}^{\prime 2}\right)}\Theta\left(\mu^2-\textbf{q}^{\prime 2}\right)\Theta\left(\textbf{q}^{\prime 2}-\mu_0^2\right)
\nonumber\\&\times \int_x^{z_M}\text{d}zP^{\left(R\right)}_{ab}\left(\alpha_s,z\right)\tilde{\mathcal{A}}_b\left(\frac{x}{z},\textbf{k}+\left(1-z\right)\textbf{q}^{\prime},\textbf{q}^{\prime 2}\right)
\label{solveme}
\end{align}
where $a$ and $b$ denote the flavour indices,  $P^{\left(R\right)}_{ab}$  are the real emission splitting kernels, $z_{M}$ is the resolution scale that separates the region for resolvable emissions from the region for non-resolvable emissions, and the Sudakov form factor $\Delta_a\left(\mu^2\right)$ gives the probability for  parton $a$ not to have 
resolvable branchings  from  scale $\mu_0$ to  scale $\mu$.   By integrating the PB evolution equation over transverse momenta, in the limit   $z_M \to 1$ one recovers  
the DGLAP~\cite{DGLAP1,DGLAP2,DGLAP3,DGLAP4} evolution equations for collinear PDFs.\\
The full solution of the PB evolution equation by iteration calculates the whole chain in the evolution containing the information about all partons and their momenta.
\section{Obtaining TMDs from final state}
In the following, we want to study how a parton shower can be used to produce effectively a transverse momentum dependent (TMD) parton density. We use Monte Carlo event generators including parton shower, determine the momentum fraction $x$, the transverse momentum $k_{\perp}$ and the scale $\mu$, and determine a parton distribution. It allows to visualize the TMD distributions and to compare them directly to TMD PDF sets. To perform the parton shower study a simple toy process is defined that allows to obtain TMD distributions as easily as possible. It contains two incoming partons $k_1$ and $k_2$ and a produced particle $q$ with the momentum $k_1+k_2=q$. One of the incoming partons $k_1$ has a fixed momentum fraction $x=0.98$ and no transverse momentum in order to make the calculation of the kinematics easier. The other parton is subject to the parton shower (or TMD distributions). The transverse momentum of $k_2$ can then be calculated from
\begin{equation}
\vec{k}_{\perp 2}=\vec{q}_{\perp}-\vec{k}_{\perp 1}
\end{equation}
This process is not a physical one but useful to investigate the parton distribution from the parton that has no fixed momentum fraction. We label it in the following as PS2TMD.\\The partonic process is generated using \textsc{pythia}8\footnote{Thanks to T. Sj\"ostrand for his help with the setup.} \cite{PYTHIA}. The events produced by \textsc{pythia} are stored in the LHE format \cite{LHE} which can be read by parton shower event generators. These events are analyzed with Rivet \cite{Rivet}, the kinematics are calculated and an effective TMD is determined. Thus a TMD is obtained from standard Monte Carlo parton showers by calculating the cumulative effect of the parton shower.\\To validate the method, TMDs are used instead of a parton shower (using the \textsc{cascade} package \cite{CASCADE,CASCADEold} (version \texttt{3.0.x})) and the results are compared to the input TMD distributions.\\For generating the events two different collinear and two different TMD PDF sets are used, PB-NLO-HERAI+II-2018-set1 and PB-NLO-HERAI+II-2018-set2 (obtained in Ref.~\cite{MartinezConnorHautmann}). They differ in the choice for the renormalization scale for the argument in $\alpha_s$. For Set1 the scale is the evolution scale $\mu^2$ and for Set2 it is the transverse momentum $q_t$.\\As a first step, TMDs are obtained from final state events using PB-TMDs to perform a consistency check.
\begin{figure}[!h]
\vspace{0.4cm}
\centering
%\hspace{-1.7cm}
\begin{minipage}[h]{0.49\textwidth}
\includegraphics[width=1.05\textwidth]{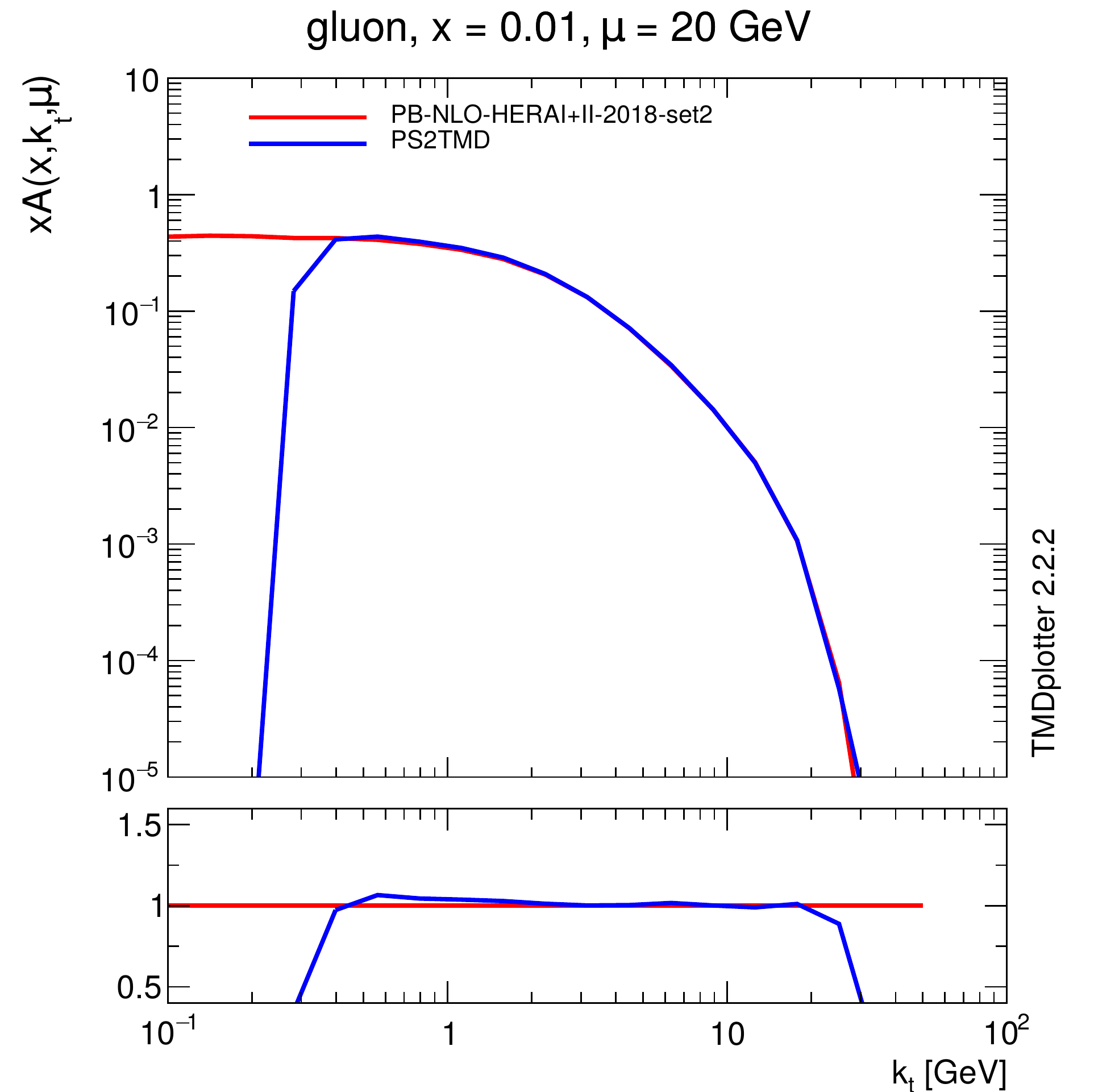}
\end{minipage}
%\hfill
%\hspace{1.7cm}
\begin{minipage}[h]{0.49\textwidth}
\includegraphics[width=1.05\textwidth]{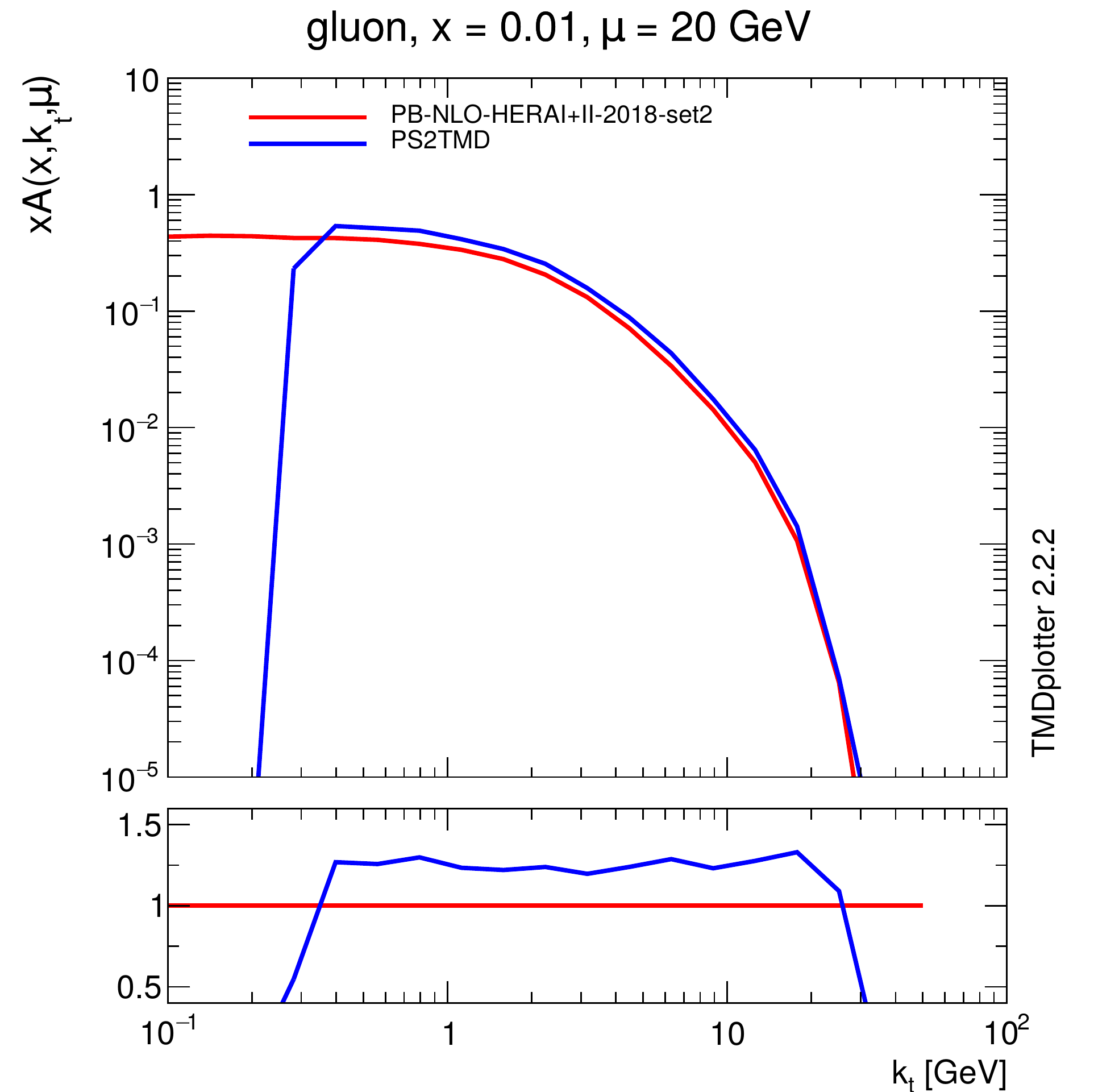}
\end{minipage}
\caption{TMD parton density as a function of $k_{\perp}$ for gluons at $\mu=20$ GeV and $x=0.01$ obtained from final state events (blue curve) and obtained from the input TMD PB-Set2 (red curve). Left: Collinear PDF: PB-Set2, TMD PDF: PB-Set2. Right: Collinear PDF: PB-Set1, TMD PDF: PB-Set2\label{PS2TMD}}
\end{figure}
\begin{figure}[!h]
\centering
%\hspace{-1.7cm}
\begin{minipage}[h]{0.49\textwidth}
\includegraphics[width=1.05\textwidth]{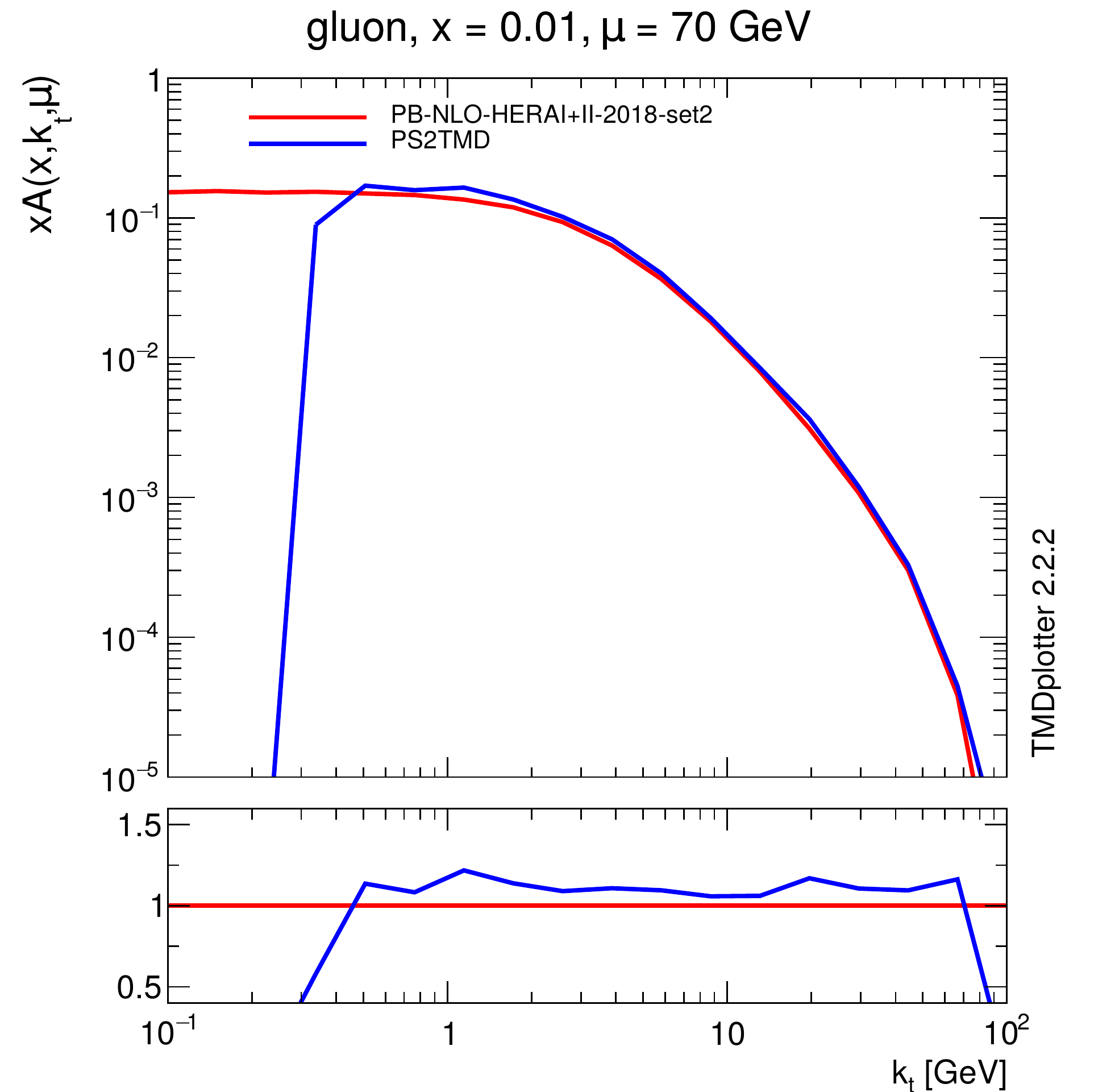}
\end{minipage}
%\hfill
%\hspace{1.7cm}
\begin{minipage}[h]{0.49\textwidth}
\includegraphics[width=1.05\textwidth]{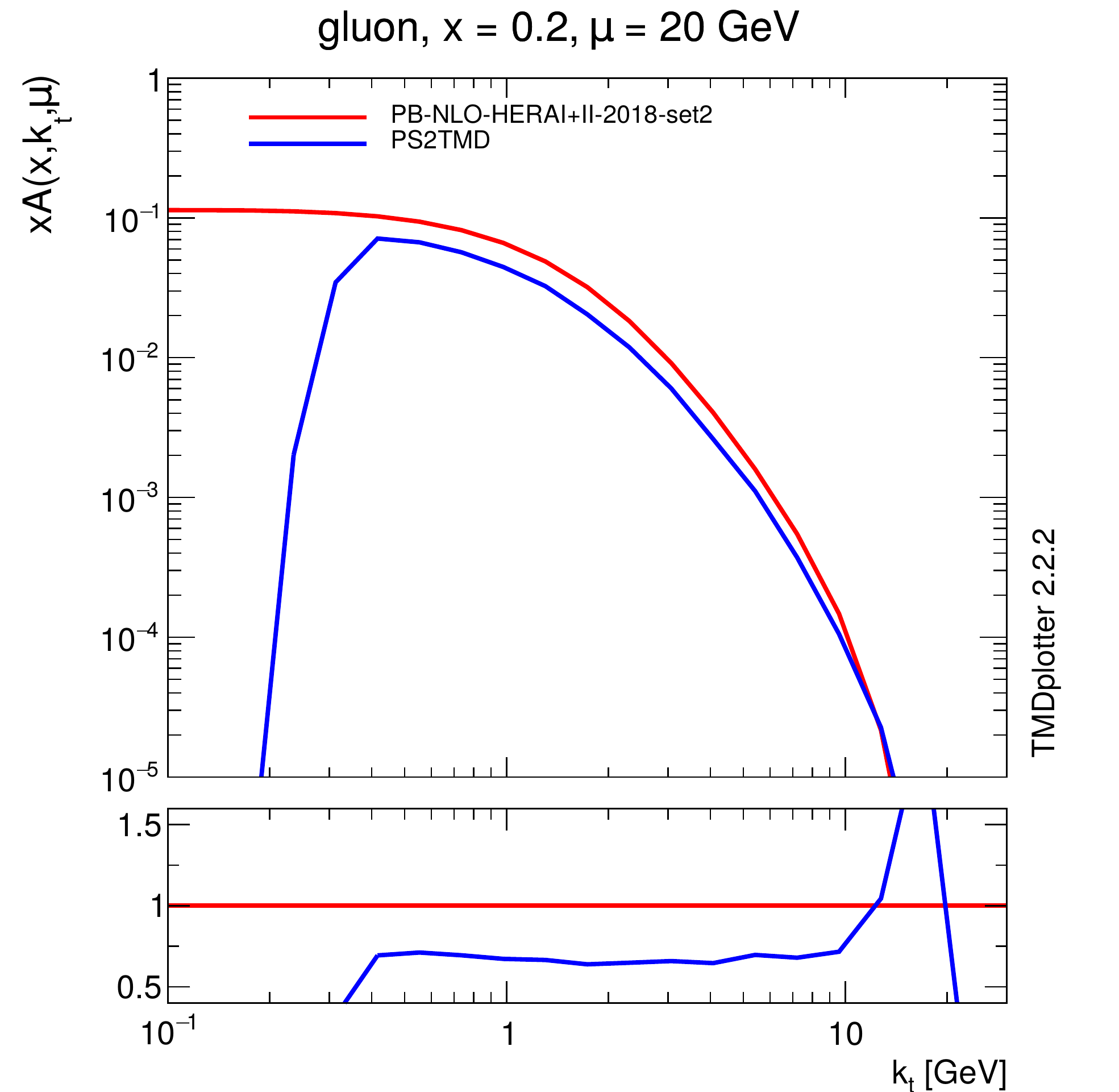}
\end{minipage}
\caption{TMD parton density as a function of $k_{\perp}$ for gluons at $\mu=70$ GeV at $x=0.01$ (left) and at $\mu=20$ GeV at $x=0.2$ (right) obtained from final state events (blue curve) and obtained from the input TMD PB-Set2 (red curve). Collinear PDF: PB-Set1, TMD PDF: PB-Set2\label{PS2TMD2}}
\end{figure}
In Fig.~\ref{PS2TMD} the TMD parton densities as a function of $k_{\perp}$ are shown for gluons at $x=0.01$ and $\mu=20$ GeV. The distribution obtained from the final state of the PS2TMD process is presented (blue curve) and compared to the PB-Set2 (red curve). In Fig.~\ref{PS2TMD} (left) the collinear PB-Set2 and the TMD PB-Set2 are used for the event generation. The distribution obtained from PS2TMD agrees well with the input distribution, showing that indeed one can obtain a TMD distribution by analysing the final state of a MC event.\\In Fig.~\ref{PS2TMD} (right) the events are generated using the collinear PB-Set1 but TMD PB-Set2. In this case a significant deviation is observed. It shows that an inconsistent use of the collinear and TMD PDF set does not reproduce the input TMD distribution. The distributions for the same configuration but at larger $\mu=70$ GeV are presented on the left in Fig.~\ref{PS2TMD2}. At this scale, the deviation between the distributions is smaller. In Fig.~\ref{PS2TMD2} (right) the distributions are shown at larger $x=0.2$ at $\mu=20$ GeV. The difference between PB-Set2 and the obtained distribution from PS2TMD is larger.\\By using TMDs for the new method PS2TMD to obtain effective TMD distributions from parton showers, the concept of the method could be validated. The comparisons show that it is essential to use the collinear and TMD PDF sets consistently.\par
\begin{figure}[!h]
\centering
%\hspace{-0.4cm}
\begin{minipage}[t]{0.49\textwidth}
\includegraphics[width=1.05\textwidth]{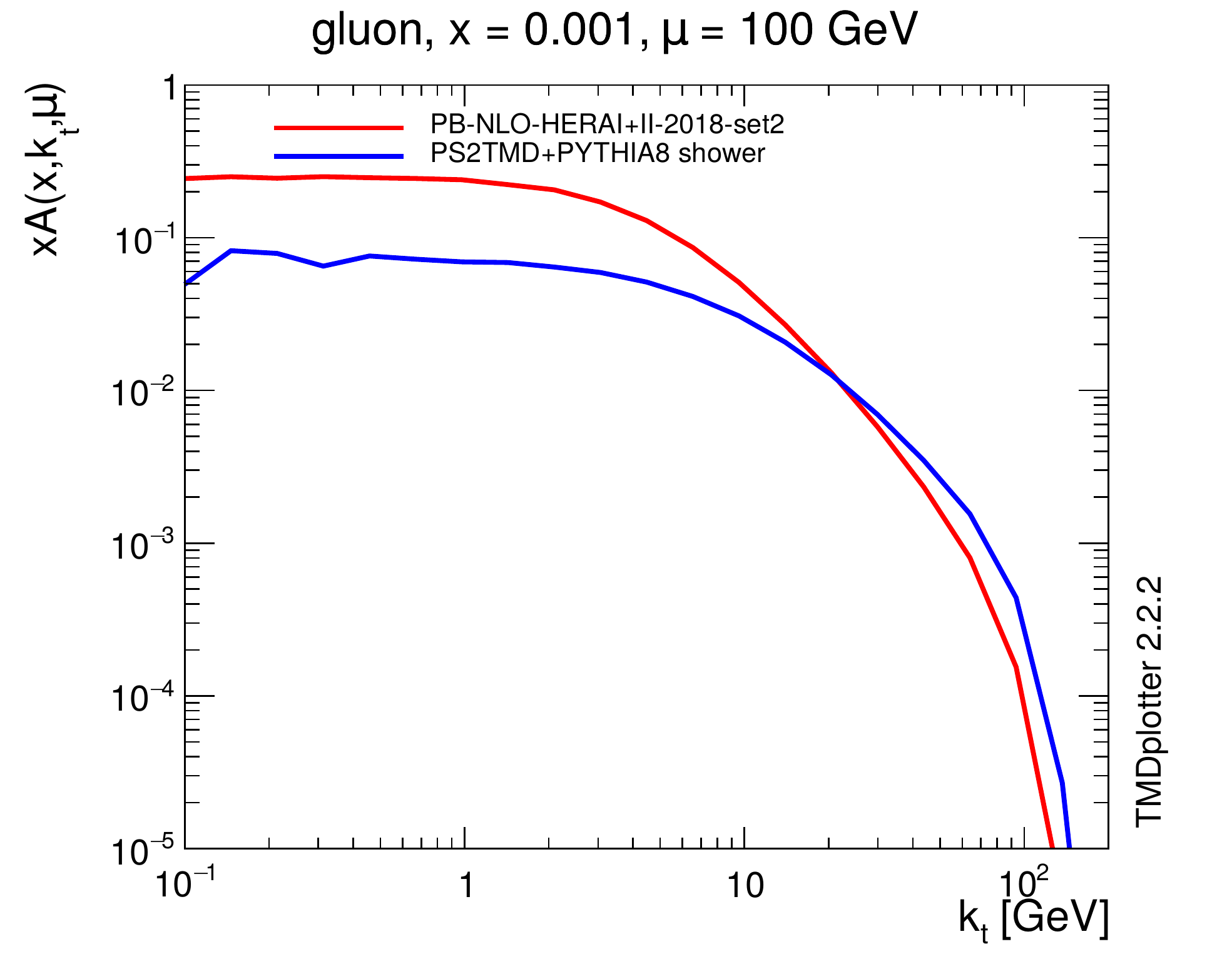}
\end{minipage}
%\hfill
%\hspace{0.6cm}
\begin{minipage}[t]{0.49\textwidth}
\includegraphics[width=1.05\textwidth]{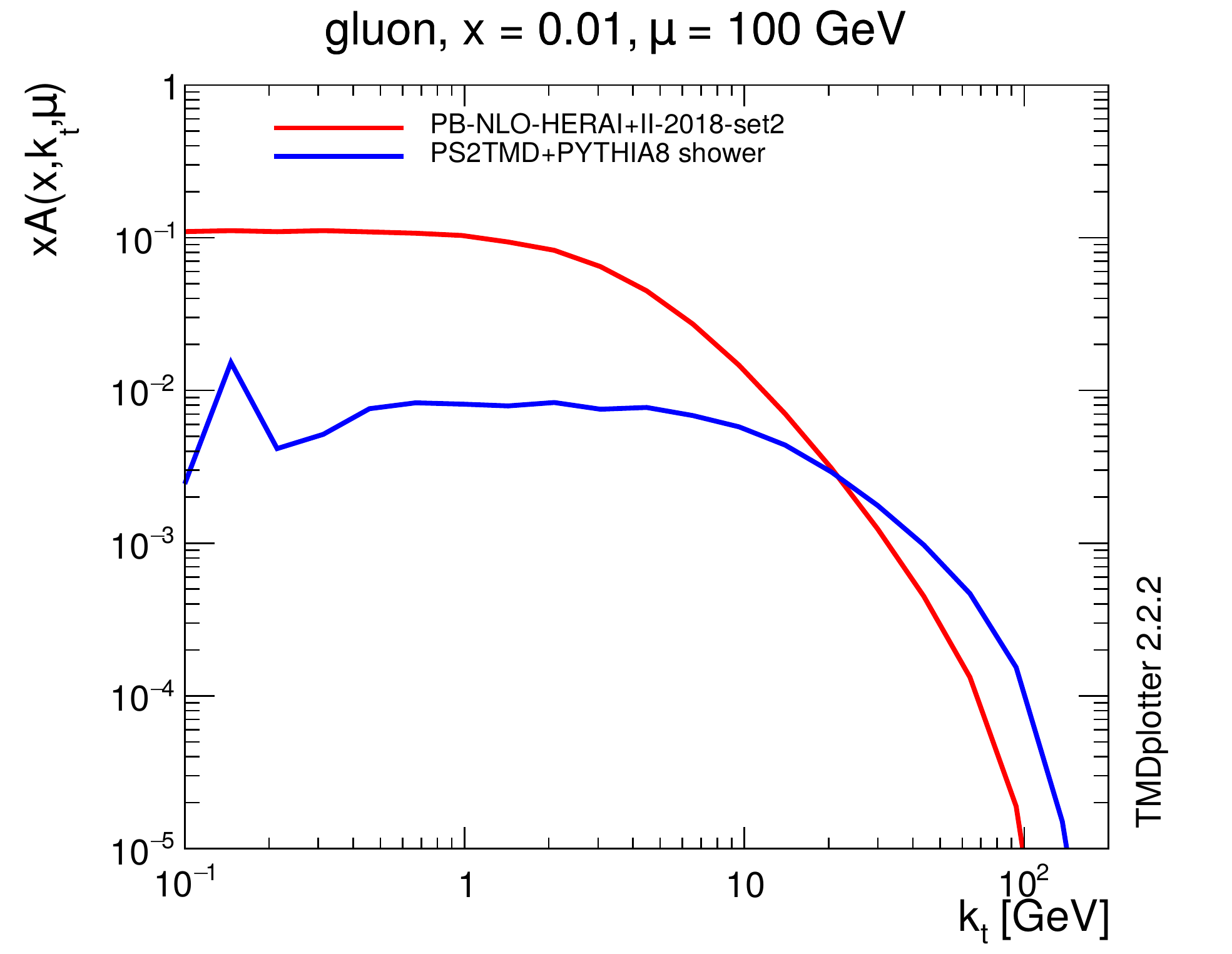}
\end{minipage}
\caption{TMD parton density as a function of $k_{\perp}$ for gluons at $\mu=100$ GeV obtained from final state events of the \textsc{pythia} shower (blue curve) and obtained from the input TMD PB-Set2 (red curve). The collinear PDF PB-Set2 is used for the generation. Left: $x=0.001$. Right: $x=0.01$.\label{PYTHIA}}
\end{figure}

In the following, effective TMDs are determined from the \textsc{pythia8} shower with the Monash 2013 tune~\cite{Monash}. \textsc{pythia} uses $p_T$-ordering for the shower. To generate the events and to produce the shower the collinear PB-Set2 is used. In Fig.~\ref{PYTHIA} the PS2TMD distributions as a function of $k_{\perp}$ obtained from the \textsc{pythia8} shower are shown. They are presented for gluons at $\mu=100$ GeV at $x=0.001$ (left) and $x=0.01$ (right), respectively, and compared to TMD PB-Set2. A large difference to the TMD PB-Set2 is observed which is more pronounced at the higher value of $x=0.01$. It has been shown explicitly in Ref.~\cite{HautmannJungLelek} that the $p_T$-ordering condition leads to flatter transverse momentum distributions. The jumps in the low-$k_{\perp}$ region come from statistical fluctuations.
\par
\begin{figure}[!h]
\centering
%\hspace{-0.4cm}
\begin{minipage}[t]{0.49\textwidth}
\includegraphics[width=1.05\textwidth]{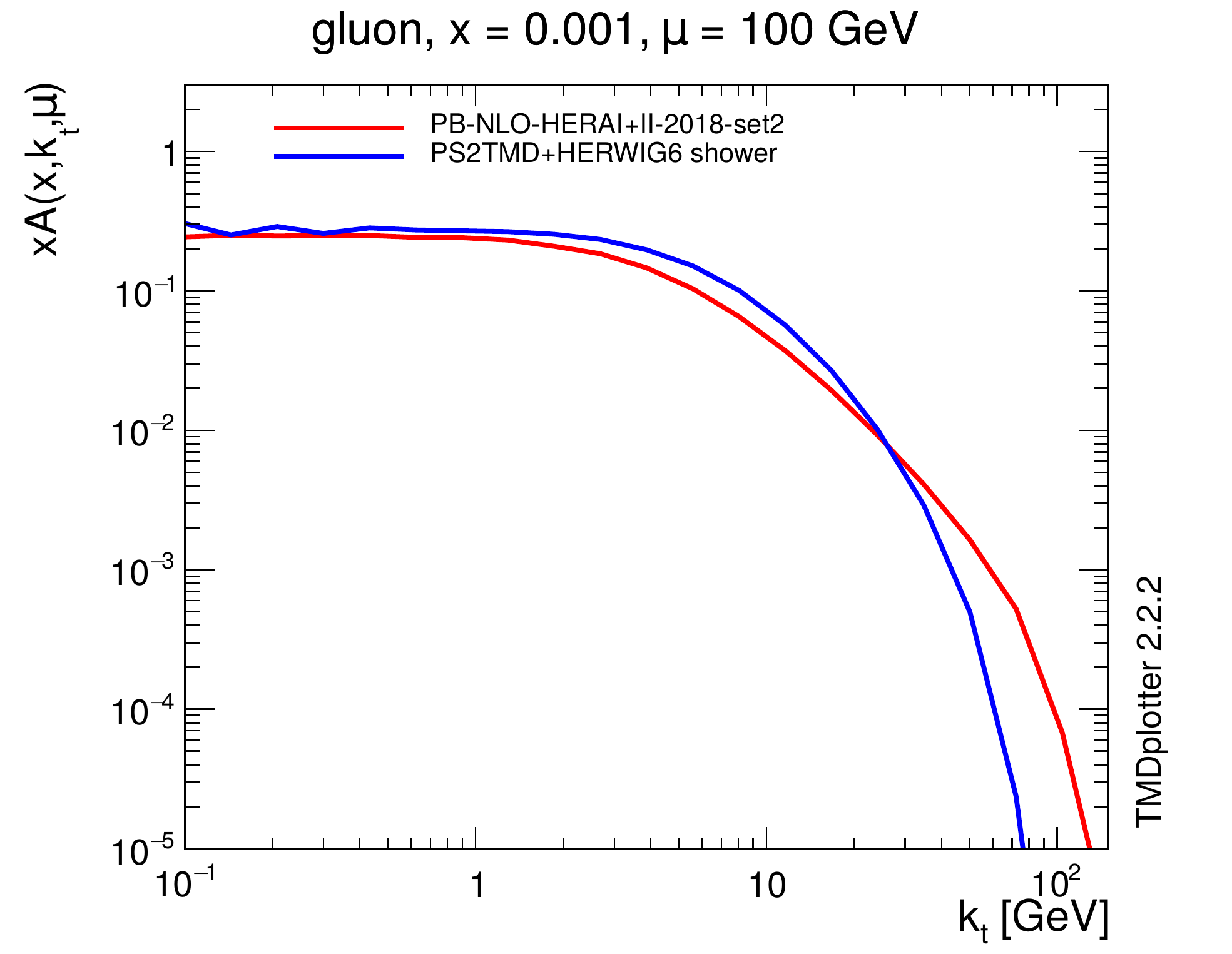}
\end{minipage}
%\hfill
%\hspace{0.6cm}
\begin{minipage}[t]{0.49\textwidth}
\includegraphics[width=1.05\textwidth]{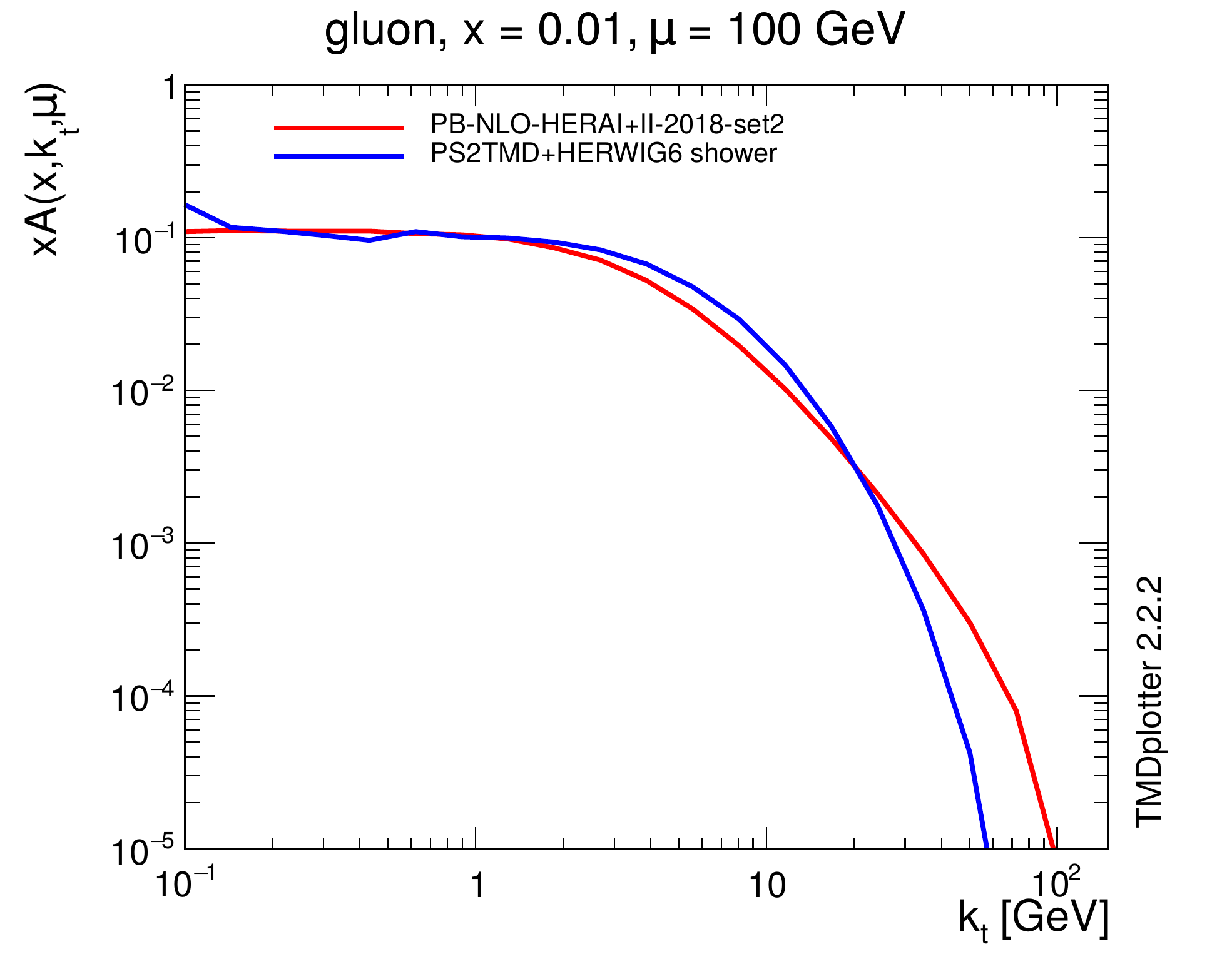}
\end{minipage}
\caption{TMD parton density as a function of $k_{\perp}$ for gluons at $\mu=100$ GeV obtained from final state events of the \textsc{herwig} shower (blue curve) and obtained from the input TMD PB-Set2 (red curve). The collinear PDF PB-Set2 is used for the generation. Left: $x=0.001$. Right: $x=0.01$.\label{herwig}}
\end{figure}

The \textsc{herwig6} parton shower with default parameter settings is studied next. It follows angular ordering. In Fig.~\ref{herwig} PS2TMDs are obtained from the \textsc{herwig6} parton shower. The parameter settings stay the same as in Fig.~\ref{PYTHIA}. On the left $x=0.001$ and on the right $x=0.01$. The collinear PB-Set2 is used for the event generation.\\
The distributions obtained from the \textsc{herwig6} shower are similar to the distributions determined from the PB-Set2 for both values of $x$. Especially for low $k_{\perp}$ they are close to each other, for larger $k_{\perp}$ they differ. 
\section{Conclusion}
A method was described to determine TMDs from final state events. The concept is proven by using TMDs, the distributions are exactly reproduced by using the collinear and TMD PDF sets consistently. The PS2TMD method is applied to the \textsc{pythia8} parton shower. Effective TMD distributions were obtained and differences to the PB-TMD distributions were observed coming from different ordering conditions in the \textsc{pythia} shower and the PB method. Effective TMD distributions were also obtained from the \textsc{herwig6} parton shower, and the distributions are closer to the PB-TMD distributions.
\section{Acknowledgements}
MS thanks DESY for giving the opportunity to go to the DIS conference. STM thanks the Humboldt Foundation for the Georg Forster research fellowship.

\end{document}